\documentclass[12pt]{article}
\def \bea{\begin{eqnarray}}
\def \beq{\begin{equation}}
\def \bra#1{\langle #1 |}
\def \eea{\end{eqnarray}}
\def \eeq{\end{equation}}

\def \ket#1{| #1 \rangle}
\def \od{\overline{D}}
\def\s{\sqrt2}
\def\st{\sqrt3}
\def\sx{\sqrt6}

\begin{document}

\rightline{TECHNION-PH-12-15}
\rightline{EFI 12-21}
\rightline{arXiv:1209.1348}

\bigskip
\centerline{\bf REVISITING $D^0$--$\od^0$ MIXING USING U-SPIN}
\bigskip
\centerline{Michael Gronau}
\centerline{\it Physics Department, Technion -- Israel Institute of Technology}
\centerline{\it Haifa 3200, Israel}
\medskip

\centerline{Jonathan L. Rosner}
\centerline{\it Enrico Fermi Institute and Department of Physics}
\centerline{\it University of Chicago, 5620 S. Ellis Avenue, Chicago, IL 60637}
\bigskip

\begin{quote}
We prove that $D^0$-$\bar D^0$ mixing in the standard model 
occurs only at second order in U-spin breaking. 
The U-spin subgroup of SU(3) is found to be a powerful tool for
analyzing the cancellation of intermediate-state contributions to
the $D^0$--$\od^0$ mixing parameter $y = \Delta \Gamma/(2 \Gamma)$.
Cancellations due to states {\it within a single U-spin triplet} 
are shown to be valid to first order in U-spin breaking.
Illustrations are given for triplets consisting of (a) pairs of
charged pions and kaons; (b) pairs of neutral pseudoscalar members of
the meson octet; (c) charged vector-pseudoscalar pairs, and (d) states
of four charged kaons and pions.
\end{quote}
\leftline{PACS numbers: 13.25.Ft, 11.30.Er, 11.30.Hv, 14.40.Lb}
\bigskip

\section{Introduction}

The parameters $x = \Delta m/\Gamma$ and $y = \Delta \Gamma/2 \Gamma$
describing mixing between $D^0$ and $\od^0$ have been established at levels of
an appreciable fraction of a percent \cite{HFAG}, 
$x=(0.63^{+0.19}_{-0.20})\%, y=(0.75 \pm 0.12)\%$. 
A key question is whether
such levels can be attained in the standard model or require new physics.

In the SU(3) limit, the contributions to $y$ of classes of intermediate
states shared by $D^0$ and $\bar D^0$ cancel one another 
in the standard model.  Previous investigations have examined the degree
to which this cancellation holds inclusively \cite{canc}.  Applying an
exclusive approach, Ref.\ \cite{Falk:2001hx} finds that for multiparticle
states near threshold, SU(3) is broken enough by phase space effects that
values of $y$ (and, generically via dispersion relations~\cite{Falk:2004wg},
$x$) of order a percent are conceivable.  This is despite the fact, proved
using the full machinery of SU(3) in Ref.\ \cite{Falk:2001hx}, that
neutral $D$ meson mixing occurs only at second order in SU(3) breaking. 

Contributions to $y$ of intermediate states with zero strangeness
typically cancel with those of states with strangeness $\pm 1$.  For example,
a contribution from the singly-Cabibbo-suppressed (SCS) transitions $D^0 \to
(\pi^+\pi^-, K^+K^-) \to \od^0$ is canceled by a contribution from the
doubly-Cabibbo-suppressed (DCS) transition $D^0 \to K^+ \pi^-$ followed by
the Cabibbo-favored (CF) transition $K^+ \pi^- \to \od^0$, plus a
contribution from the CF transition $D^0 \to K^- \pi^+$ followed by the DCS
transition $K^- \pi^+ \to \od^0$.  The intermediate states in this case
comprise a single triplet of the SU(3) subgroup known as U-spin \cite{Usp}.
Just as the fundamental representation of I-spin (isospin) is composed of
$(u,d)$, that of U-spin is $(d,s)$.

U-spin symmetry has been known for a long time to provide useful relations
among amplitudes of hadronic $D$ decays~\cite{Kingsley,Gronau:2000ru}.
Typical U-spin breaking, described by quantities such as
$(m_s - m_d)/\Lambda_{\rm QCD}$ or $f_K/f_\pi-1$, is of order $0.2-0.3$ and
may be treated perturbatively in hadronic matrix elements.
A very early suggestion was made in Ref.\,\cite{Savage:1991wu}
that SU(3) breaking at this level in a penguin amplitude may account for the 
somewhat unexpected large value of the ratio of branching ratios 
${\cal B}(D^0\to K^+K^-)/{\cal B}(D^0\to\pi^+\pi^-)=2.8$\,\cite{PDG}.
A recent study of $D$ decays into two 
pseudoscalars\,\cite{Bhattacharya:2012ah,Bhattacharya:2012kq} 
has shown that U-spin breaking  at a level between 10 to 20 percent 
in an enhanced  
nonperturbative penguin amplitude may account well for 
this ratio and for the unexpected large difference 
between CP asymmetries measured recently in these two 
processes~\cite{Aaij:2011in,CDF}. Two other  
studies discussing these two effects of U-spin breaking 
have been presented recently in Refs.\,\cite{Feldmann:2012js}
and~\cite{Brod:2012ud}.

In this paper we shall show that a cancellation, to first order in U-spin 
breaking, of contributions to $D^0$--$\od^0$ mixing within single U-spin
triplets is a very general result.  In Sec.\ II we apply U-spin and its
breaking to a $D^0$--$\od^0$ mixing amplitude, $A_{D \bar D}\equiv 
\langle \bar D^0|H_W\,H_W|D^0\rangle$.  We relate this transition
amplitude to $\Delta \Gamma$ in Sec.\ III,
and derive a general U-spin sum rule corresponding to the cancellation
of contributions to $\Delta \Gamma$ in Sec.\ IV.  Examples of these sum rules
for pairs of charged pseudoscalar mesons, pairs of neutral pseudoscalar
mesons, and charged vector-pseudoscalar pairs, are given in Sec.\ V.
Some results are noted in Sec.\ VI for states of four charged
pions and kaons, while Sec.\ VII concludes.

\section{U-spin breaking in a $D^0$--$\bar D^0$ mixing amplitude}\label{sec:ADD}

Let us consider an amplitude which connects $D^0$ and $\bar D^0$ 
through second order weak interactions,
\beq\label{ADD}
A_{D\bar D} \equiv \langle \bar D^0|H_W^{\Delta C=-1}\,H_W^{\Delta C=-1}|D^0\rangle~.
\eeq
We will now show that this amplitude vanishes in the U-spin symmetry limit, and
that it vanishes also when including first order U-spin corrections.
Keeping the flavor structure of the $\Delta C=-1$ weak Hamiltonian but 
suppressing its Lorentz structure and denoting 
$C\equiv \cos\theta_c, S \equiv \sin\theta_c$, one has
\beq
H_W^{\Delta C=-1} = 
\frac{G_F}{\sqrt 2}
[C(\bar s c) - S(\bar d c)][C(\bar u d) + S(\bar u s)]~.
\eeq
Only six out of the sixteen terms in $H_WH_W$ obey $\Delta S=0$ and 
contribute to $A_{D\bar D}$:
\bea\label{6terms}
A_{D\bar D} & = & \frac{G^2_F\,C^2S^2}{2}\langle \bar D^0|- [(\bar d c)(\bar u s)][(\bar s c)(\bar u d) 
-[(\bar s c)(\bar u d)][(\bar d c)(\bar u s)]+ [(\bar d c)(\bar u d)][(\bar d c)(\bar u d)] \nonumber\\
& &~~~~~~~~ + [(\bar s c)(\bar u s)][(\bar s c)(\bar u s)] 
- [(\bar d c)(\bar u d)][(\bar s c)(\bar u s)] - [(\bar s c)(\bar u s)][(\bar d c)(\bar u d)]|D^0\rangle
\eea

We will now show that the operator contributing to $A_{D\bar D}$ in Eq.(3) transforms like $U=2, U_3=0$. Neglecting $c \bar u$ terms, which are singlets under U-spin, this operator reduces to
\beq
{\cal O}_{D\bar D} = -(\bar d s)(\bar s d) -  (\bar s d)(\bar d s)
+[(\bar s s) - (\bar d d)][(\bar s s) - (\bar d d)]~.
\eeq
Taking the quark and antiquark pairs of states, $(|d\rangle, |s\rangle)$ 
and $(|\bar s\rangle, -|\bar d\rangle)$, to be U-spin-doublets, we find the 
following behavior of $\bar q q'$ operators under U-spin,
\beq\label{qbarq}
(\bar s d)  =  (1, -1)~,~~~~
-(\bar ds) =  (1, 1)~,~~~~
(\bar ss) - (\bar dd) = \sqrt2(1,0)~,
\eeq
implying
\beq
{\cal O}_{D\bar D} = (1, 1)\otimes(1, -1) + (1, -1)\otimes (1, 1) + 2(1, 0)\otimes (1, 0)~.
\eeq
Using Clebsch-Gordan coefficients for $1\otimes 1$, the operator ${\cal O}_{D\bar D}$ is seen to transform  as pure $U=2, U_3=0$.

$D^0$ and $\bar D^0$ are U-spin singlets. Therefore $A_{D\bar D}$ vanishes  in the U-spin symmetry limit. Assuming that U-spin breaking may be treated perturbatively, a U-spin breaking mass term ($\propto \bar ss - \bar dd$) behaves like $U=1, U_3=0$.  Thus $A_{D\bar D}$ vanishes also in the presence of first order U-spin-breaking corrections,
and may obtain a nonzero value only when including  second order U-spin breaking. For a short notation, we will refer to this behavior of vanishing in the U-spin symmetry limit including first order U-spin breaking corrections as {\em vanishing in USFB}.  

Ref.\,\cite{Falk:2001hx} presented a lengthy SU(3) group theoretical  argument involving high representations of this group showing
that $A_{D\bar D}$ vanishes in the limit of flavor SU(3) symmetry and 
when including first order SU(3) breaking corrections. We have shown 
that this behavior is actually due to only U-spin, a particular SU(2) subgroup of SU(3). 

\section{Relation between $A_{D\bar D}$ and the width difference $\Delta\Gamma$}\label{sec:DeltaGamma}

Let us consider the width-difference between the two neutral $D$  mass eigenstates, neglecting CP violation in $D^0$-$\bar D^0$ 
mixing~\cite{Lee:1957qq,Branco:1999fs,Bigi:2000yz} 
\beq\label{DeltaGamma}
\Delta\Gamma = \sum\limits_{f^D}
\,\rho(f^D)\,\left(\langle \bar D^0|H_W^{\Delta C=-1}|f^D\rangle
\langle f^D|H_W^{\Delta C=-1}|D^0\rangle + c.c.\right)~.
\eeq
Here $|f^D\rangle$ are normalized states into which $D^0$ and $\bar D^0$ 
may decay, while $\rho(f^D)$ are corresponding densities of states, i.e., phase space factors. 

The entire set of final states $|f^D\rangle$ accessible to $D^0$ 
decays, also accessible to $\bar D^0$ decays, may be divided into three subsets according to their strangeness $S$. Each of these three subsets, 
$\{|f^D_S\rangle\}$, $S=-1, 0, 1$, consisting  of states with
an arbitrary number of final hadrons, is complete in a quantum mechanics sense. That is, an arbitrary final state with strangeness $S$ in $D^0$ 
decay may be expressed as a linear superposition of states  belonging to this set. Thus, each of the above three subsets obeys a completeness relation,
 \beq\label{completeness}
\sum\limits_{f^D_S}\,\rho(f^D_S)\,|f^D_S\rangle\langle f^D_S | = 1~,~~~(S=-1, 0, +1)~.
\eeq

The $\Delta C=-1$ weak Hamiltonian involves three parts, $H_{\Delta S}$, corresponding to $\Delta S=-1, 0, 1$,
\beq
H_W^{\Delta C=-1}  = H_{-1} + H_{0} + H_{+1}~.
\eeq
The three operators on the right-hand-side, behaving like three components, $U_3=-1,0,+1$, of a U-spin vector, have the following quark flavor structure
\beq\label{H_S}
H_{-1} = \frac{G_F\,C^2}{\sqrt 2}\,(\bar s c)(\bar u d)\,,~~~~
H_0 = \frac{G_F\,CS}{\sqrt 2}\,[(\bar s c)(\bar u s) - (\bar d c)(\bar u d)]\,,~~~~
H_{+1} = -\frac{G_F\,S^2}{\sqrt 2}\,(\bar d c)(\bar u s)~.
\eeq
The sum in (\ref{DeltaGamma}) may thus be divided into three sums over the three independent complete sets of normalized states for a given strangeness,
\beq
\Delta\Gamma = 
\sum\limits_{S=\mp1,0}\sum\limits_{f^D_S}\,\rho(f^D_S)\langle \bar D^0|
H_{-S}|f^D_S\rangle\langle f^D_S|H_S|D^0\rangle + c.c.~.
\eeq  
Using (\ref{completeness}) one has
\beq
\Delta\Gamma  =  \langle \bar D^0|H_{+1}\,H_{-1} + H_{-1}\,H_{+1}
+ H_0\,H_0|D^0\rangle +  c.c.~.
\eeq
Thus one finds 
\beq
\Delta\Gamma = 
\langle \bar D^0|H_W^{\Delta C=-1}\,H_W^{\Delta C=-1}|D^0\rangle + 
c.c.~,
\eeq
implying
\beq\label{y-Delta}
\Delta\Gamma =  2{\rm Re}(A_{D\bar D})~,
\eeq
where $A_{D\bar D}$ has been defined in (\ref{ADD}).

As far as we are aware, this relation between $A_{D\bar D}$ 
and the width difference $\Delta\Gamma$ has not been
shown explicitly in the literature. See, for instance, discussions in 
Refs.
\cite{Falk:2001hx,Branco:1999fs,Bigi:2000yz,Bigi:2000wn,Burdman:2003rs} 
and in numerous other references quoted in these three papers and two 
books. An implicit dependence of $\Delta\Gamma$ on $A_{D\bar D}$ has 
been assumed 
in Ref.\,\cite{Falk:2001hx} without discussing any details or quantitative arguments for this dependence. The 
dependence found in Eq.\,(\ref{y-Delta}) is extremely simple.

In the previous section we have shown that $A_{D\bar D}$ vanishes in 
USFB. Eq.\,(\ref{y-Delta}) implies that this same U-spin behavior applies 
also to $\Delta\Gamma$ and therefore to the dimensionless mixing
parameter $y\equiv \Delta\Gamma/2\Gamma = {\rm Re}(A_{D\bar D})/\Gamma$.

\section{A general U-spin sum rule  $\Delta\Gamma({\{f_{U=1}\}})=0$}

In the preceding two sections we have shown that $\Delta\Gamma$ 
in (\ref{DeltaGamma}) vanishes in USFB when summing over all 
possible $D^0$ decay final states. We now wish to identify 
subsets of these states $\{f^D_i\}$ for which 
$\Delta\Gamma(\{f^D_i\})=0$ holds in this same approximation. For 
this purpose we search for subsets of 
states for which the following relation similar to (\ref{y-Delta})
holds in USFB,
\beq
\Delta\Gamma(\{f^D_i\}) \propto  2{\rm Re}(A_{D\bar D})~.
\eeq

We have seen that the three terms of given strangeness in 
$H_W^{\Delta C=-1}$ transform as three components of a $U=1$ 
operator. Consequently only intermediate states  with $U = 1$ 
contribute to $\Delta\Gamma$. 
We choose our subset of states to be three components of a 
U-spin triplet, $\{f_{U=1,U_3}\,;~U_3= 0, \pm 1\}$,
\beq\label{U=1Sum}
\Delta\Gamma(\{f_{U=1}\}) = \langle \bar D^0|H_W^{\Delta C=-1}
(\sum\limits_{U_3=0,\pm 1}\rho(f_{U=1,U_3})|f_{U=1,U_3}
\rangle\langle f_{U=1,U_3}|)\,H_W^{\Delta C=-1}|D^0\rangle + c.c.
\eeq
Assuming for a moment U-spin symmetric phase space factors, one has
\beq\label{U=1completeness}
\sum\limits_{U_3=\pm 1,0}\rho(f_{U=1,U_3}|f_{U=1,U_3}
\rangle\langle f_{U=1,U_3} | = 
\rho(f_{U=1})\sum\limits_{U_3=\pm 1,0}|f_{U=1,U_3}
\rangle\langle f_{U=1,U_3} | = \rho(f_{U=1})\,{\bf 1} 
\eeq
implying
\beq\label{Delta}
\Delta\Gamma(\{f_{U=1}\})  = 
\rho(f_{U=1})\langle \bar D^0|H_W^{\Delta C=-1}
H_W^{\Delta C=-1}|D^0\rangle + c.c. = 0~.
\eeq
This result holds also in the presence of first order U-spin breaking in  
$\langle f_{U=1}|H_W|D^0\rangle$.and $\langle \bar D^0|H_W|f_{U=1}\rangle$
as such corrections behaving like $U=1, U_3=0$ cancel in (\ref{Delta}). 
First order U-spin breaking in phase space factors may be made to cancel by
a judicious choice of low mass final states. This will be 
demonstrated below in several specific examples.

Neglecting CP violation in $D^0$ decays and denoting 
$|\bar f \rangle \equiv CP|f\rangle$, one has
\beq
\langle \bar D^0|H_W|f\rangle = \langle \bar f|H_W|D^0\rangle^*~,
\eeq
implying
\beq
\langle \bar D^0|H_W|f\rangle \langle f|H_W|D^0\rangle + c.c. 
= 2{\rm Re}(\langle \bar f|H_W|D^0\rangle^*\langle f|H_W|D^0\rangle)~.
\eeq
The generic form of a U-spin sum rule which holds in USFB 
by a judicious choice of final states is thus
\beq\label{USR}
{\rm Re}\,[\hskip-3mm\sum\limits_{~~U_3=\pm 1,0}\langle \bar f_{1,U_3}|H_W|
D^0\rangle^*\langle f_{1,U_3}|H_W|D^0\rangle] = 0~.
\eeq
We note that the states $|f\rangle$ and $|\bar f\rangle$ do not necessarily 
belong to the same $U=1$ representation. For instance $|K^{*+}\pi^-\rangle$ 
and $|K^{*-}\pi^+\rangle$, which are each other's CP-conjugates, are $U_3=1$ 
and $U_3=-1$ states in two different $U=1$ triplets.

In the special case that $|f_{1,0}\rangle$ is a CP eigenstate with  
eigenvalue $\eta_{CP}$ we will denote
$CP|f_{1,U_3}\rangle=\eta_{CP}|\bar f_{1,U_3}\rangle$ for
all three triplet states. [Note that while $|\bar f_{1,U_3}\rangle$
is a state transforming as  $|1,-U_3\rangle$, the two states 
$|\bar f_{1,1}\rangle$ and $|f_{1,-1}\rangle$ may differ by a sign. 
See Eq.\,(\ref{piK}) below.] Using this convention one finds
\bea\label{U3=0}
\langle \bar D^0|H_W|f_{1,0}\rangle\,\langle f_{1,0}|H_W|D^0\rangle + c.c. 
& = & 2\eta_{CP}|\langle f_{1,0}|H_W|D^0\rangle|^2~,\\
\langle \bar D^0|H_W|f_{1,\pm 1}\rangle\,\langle f_{1,\pm 1}|H_W|
D^0\rangle + c.c. 
& = & 2\eta_{CP}{\rm Re}(\langle \bar f_{1,1}|H_W|D^0\rangle^*
\langle f_{1,1}|H_W|D^0\rangle)~.
\eea
Thus we have derived the following generic form for a U-spin sum rule in USFB for
triplet states of which $|f_{1,0}\rangle$ is a CP eigenstate, 
\beq\label{USRcp}
\frac{\eta_{CP}}{2}\Delta\Gamma(\{f_{U=1}\}) = |\langle f_{1,0}|H_W|D^0\rangle|^2
+ 2{\rm Re}(\langle \bar f_{1,1}|H_W|D^0\rangle^*\,\langle f_{1,1}|H_W|D^0\rangle) = 0~.
\eeq
Note that the triplet states $|f_{1,U_3}\rangle$ may be admixtures of low mass
physical states.  We will now demonstrate the sum rule (\ref{USRcp}) and
corresponding expressions for $y(\{f_{U=1}\})$ in several  examples.

\section{Examples of $U=1$ sum rules}

\subsection{$D^0$ decays to pairs of charged pseudoscalar mesons, 
$\pi^\pm, K^\pm$}

The pairs $(\pi^-,K^-)$ and $(K^+,-\pi^+)$ are U-spin doublets.
The four possible two-particle states can be written in the form 
of U-spin states:
\beq\label{piK}
|\pi^-K^+\rangle = |1, 1\rangle\,,~~~~
|K^-\pi^+\rangle = -|1,-1\rangle\,,~~~~
\eeq
\beq
\frac{1}{\s} |K^-K^+ - \pi^-\pi^+ \rangle = |1,0\rangle\,~~~~
\frac{1}{\s} |K^-K^+ + \pi^-\pi^+ \rangle = |0,0\rangle~.
\eeq
Using $\langle 0,0|H_W|D^0\rangle = 0$ one may write
\bea
|\langle 1,0|H_W|D^0\rangle|^2 & = & |\langle 1,0|H_W|D^0\rangle|^2
+ |\langle 0,0|H_W|D^0\rangle|^2 
\nonumber\\
& = & |\langle K^-K^+|H_W|D^0\rangle|^2 + |\langle \pi^-\pi^+|H_W|D^0\rangle|^2~.
\eea
Consequently the sum rule (\ref{USR}) reads
\bea
\frac{1}{2}\Delta\Gamma(\pi^\pm, K^\pm) & = & 
|\langle K^-K^+|H_W|D^0\rangle|^2 + |\langle \pi^-\pi^+|H_W|D^0\rangle|^2
\nonumber\\
& + &2{\rm Re}(\langle \pi^-K^+|H_W|D^0\rangle^*\,\langle K^-\pi^+|H_W|D^0\rangle) = 0~.
\eea

A corresponding expression for 
$y(\pi^\pm, K^\pm)=\Delta\Gamma(\pi^\pm, K^\pm)/2\Gamma$  is 
obtained in terms of branching ratios and the strong phase difference 
$\delta$ between amplitudes for $D^0\to \pi^-K^+$ and 
$D^0\to K^-\pi^+$,
\beq\label{y+-}
y(\pi^\pm, K^\pm) = {\cal B}(D^0 \to \pi^-\pi^+) + {\cal B}(D^0 \to K^-K^+) - 
2\cos\delta\sqrt{{\cal B}(D^0 \to K^-\pi^+){\cal B}(D^0 \to \pi^-K^+)}~.
\eeq
The minus sign of the last term on the right-hand-side may be traced 
back  to three minus signs appearing in the second operator equation 
(\ref{qbarq}), the third operator equation (\ref{H_S}) and the second 
state equation (\ref{piK}). 
 The strong phase difference $\delta$ vanishes in the 
 U-spin symmetry limit~\cite{Wolfenstein:1995kv}, and $\cos\delta=1$ holds
 up to a first order U-spin breaking correction. Thus the quantity 
 $y(\pi^\pm, K^\pm)$ which vanishes in USFB is given by
 \beq \label{eqn:chpk}
y(\pi^\pm, K^\pm) = {\cal B}(D^0 \to \pi^-\pi^+) + {\cal B}(D^0 \to K^-K^+) - 
2\sqrt{{\cal B}(D^0 \to K^-\pi^+){\cal B}(D^0 \to \pi^-K^+)}=0~.
\eeq  

Using updated branching fractions \cite{PDG}
\bea
{\cal B}(D^0 \to \pi^+ \pi^-) & = & (1.401 \pm 0.027) \times 10^{-3}~,\\
{\cal B}(D^0 \to K^+ K^-) & = & (3.96 \pm 0.08) \times 10^{-3}~,\\
{\cal B}(D^0 \to K^- \pi^+) & = & (3.88 \pm 0.05)\%~,\\
{\cal B}(D^0 \to K^+ \pi^-) & = & (1.31 \pm 0.08) \times 10^{-4}~,
\eea
we find $y(\pi^\pm, K^\pm) = (0.85 \pm 0.17)\times 10^{-3}$, 
substantial cancellation between positive and negative terms in
(\ref{eqn:chpk}) (corresponding to almost an order of magnitude
suppression),  and an order of magnitude below the observed value of
$y = (0.75 \pm 0.12$)\% (no CP violation assumed \cite{HFAG}).

One can see that U-spin breaking cancels to first order in phase space factors
in Eq.\ (\ref{eqn:chpk}).  
Expand the ratios of phase space factors $\rho(K^+K^-)/\rho(\pi^+\pi^-)$ 
and $\rho(\pi^\pm K^\mp)/\rho(\pi^+\pi^-)$  to first order in 
$\Delta \equiv (m^2_K-m^2_\pi)/m^2_D$ [equivalently, to first order in
$(m_s-m_d)/m_c$]. The coefficients of
$\Delta$ in the three terms on the right-hand side of Eq.\ (\ref{eqn:chpk}) 
are in the ratio $0:-2:2$ and hence cancel one another.

Ref.\,\cite{Falk:2001hx} discussed contributions to $y$ from 
two pseudoscalar states belonging to a common SU(3) representation.
The expression (\ref{y+-}) corresponding to states which are members of 
a U-spin triplet has been considered as an arbitrary partial contribution 
to this value of $y$, without motivating that choice and without noticing that 
$y(\pi^\pm,K^\pm)$ in Eq.\,(\ref{eqn:chpk}) vanishes to first order U-spin breaking.

\subsection{Decays to pairs of neutral pseudoscalar mesons, 
$\pi^0, \eta, K^0, \bar K^0$}

When considering final states involving two neutral pseudoscalar
mesons we will neglect $\eta-\eta'$ mixing by taking $\eta = \eta_8$.
This approximation does not spoil the derived USFB sum rule 
because $\eta-\eta'$ mixing is due to first order U-spin breaking 
transforming as $U=1$. 

The following superpositions of single-particle states belong to
a U-spin triplet:
\beq\label{singleU=1}
|K^0\rangle = |1,1\rangle\,,~~~~~|\bar K^0\rangle =  - |1, -1\rangle\,,~~~~~
\frac{1}{2}(\sqrt3\,|\eta\rangle - |\pi^0\rangle) = |1, 0\rangle~,
\eeq  
while the orthogonal U-spin singlet is
\beq\label{singleU=0}
\frac{1}{2}(|\eta\rangle + \sqrt3\,|\pi^0\rangle) = |0, 0\rangle~.
\eeq
Here we have used the convention $\pi^0 = (d \bar d - u \bar u)/\s$,
$\eta = (2 s \bar s - u \bar u - d \bar d)/\sx$, and all states are labeled
by $\ket{U,U_3}$.

We now form U-spin multiplets out of pairs of the above states.  
Consider first the states with $U_3 = 1$:
\bea
\ket{2,1} & = & (\ket{1,1} \otimes \ket{1,0} + \ket{1,0} \otimes
\ket{1,1})/\s~, \\
\ket{1,1} & = & (\ket{1,1} \otimes \ket{1,0} - \ket{1,0} \otimes
\ket{1,1})/\s~, \label {eqn:anti1} \\
\ket{1',1} & = & \ket{1,1} \otimes \ket{0,0}~,
\eea
where the states on the left are two-particle states, while those on the
right are one-particle states given in Eqs.\,(\ref{singleU=1}) and
(\ref{singleU=0}) in terms of neutral pseudoscalar mesons. By Bose statistics
we need not consider the state (\ref{eqn:anti1}) as it is made of an
antisymmetric product.  We shall also need the two-particle states
with $U_3 = 0$.  There are two $U=0$ states
\bea
\ket{0',0} & = & \ket{0,0} \otimes \ket{0,0}~, \\
\ket{0,0}  & = & (\ket{1,1} \otimes \ket{1,- \! 1} + \ket{1,- \! 1} \otimes
\ket{1,1} - \ket{1,0} \otimes \ket{1,0})/\st~,
\eea
two $U=1$ states
\bea
\ket{1',0} & = & \ket{1,0} \otimes \ket{0,0}~, \\
\ket{1,0}  & = & (\ket{1,1}  \otimes \ket{1,- \! 1} - \ket{1,- \!1 } \otimes
\ket{1,1})/\s \label{eqn:anti0}~,
\eea
and one $U=2$ state
\beq
\ket{2,0} = (\ket{1,1} \otimes \ket{1,- \! 1} + \ket{1,- \! 1} \otimes
\ket{1,1} + 2 \ket{1,0} \otimes \ket{1,0})/\sx~.
\eeq
Again, by Bose statistics, we need not consider the state (\ref{eqn:anti0})
further.  Now we calculate the contribution to $y$ of decay amplitudes
participating in the transition $D^0 \to \od^0$ due to pairs of neutral
mesons belonging to the $U=1$ multiplet.  We first discuss the contributions
of the $S=\pm 1$ states $K^0 \pi^0, K^0 \eta^0, \bar K^0 \pi^0$, and
$\bar K^0 \eta$.

As $H_W$ transforms according to $\Delta U = 1$, and the initial $D^0$
has $U=0$, the transition amplitude $\bra{2,1} H_W \ket{D^0}$ vanishes.
Expressed in terms of physical mesons, this means
\beq \label{eqn:U21}
[\st A(D^0 \to K^0 \eta) - A(D^0 \to K^0 \pi^0)]/2 = 0~.
\eeq
We also have the transition of interest,
\beq
\bra{1',1} H_W \ket{D^0} = [A(D^0 \to K^0 \eta) + \st A(D^0 \to K^0 \pi^0)]/2
= (2/\st)A(D^0 \to K^0 \pi^0)~,
\eeq
where (\ref{eqn:U21}) was used in the second equality. 
Thus, in analogy with the last term in Eq.\ (\ref{y+-}), one gets
a contribution to $y$ of the form
\beq
y (|\Delta S| = 1) = -(8/3)\sqrt{{\cal B}(D^0 \to K^0\pi^0)
{\cal B}(D^0 \to \bar K^0 \pi^0)}~.
\eeq
Using (\ref{eqn:U21}) one obtains a contribution to $y$ from the 
$\Delta S = \pm 1$ transitions involving all pairs of neutral octet
members,
\beq
y (|\Delta S| = 1) = - 2 \sqrt{[{\cal B}(D^0 \to K^0 \eta) + {\cal B}(D^0 \to K^0
\pi^0)] [{\cal B}(D^0 \to \bar K^0 \eta) + {\cal B}(D^0 \to \bar K^0 \pi^0)]}~.
\eeq
Here, as in the case of charged pions and kaons, one may
neglect the cosine of  a strong phase difference which is second 
order in U-spin breaking.

Now we turn to the SCS ($\Delta S = 0$) transitions $\bra{1',0} H_W
\ket{D^0}$.  We have a number of relations between amplitudes for
$D^0$ decays to $\eta \eta$, $\eta \pi^0$, and $\pi^0 \pi^0$, and will
find the usual SU(3) result $A(D^0 \to K^0 \bar K^0) = 0$, stemming
from the vanishing of the transitions $\bra{0',0} H_W \ket{D^0}$,
$\bra{0,0} H_W \ket{D^0}$, and $\bra{2,0} H_W \ket{D^0}$.  The first
of these implies
\beq \label{eqn:0',0}
A(D^0 \to \eta \eta) + 2 \st A(D^0 \to \eta \pi^0) + 3 A(D^0 \to \pi^0
\pi^0) = 0~.
\eeq
Linear combinations of the second and third imply $A(D^0 \to K^0 \bar K^0)
= 0$ and the relation
\beq \label{eqn:20,0}
3 A(D^0 \to \eta \eta) - 2 \st A(D^0 \to \eta \pi^0) + A(D^0 \to \pi^0
\pi^0) = 0~.
\eeq

The transition of interest is
\beq \label{eqn:1',0}
\bra{1',0} H_W \ket{D^0} = \frac{\st}{4} A(D^0 \to \eta \eta) + \frac{1}{2}
A(D^0 \to \eta \pi^0) - \frac{\st}{4} A(D^0 \to \pi^0 \pi^0)~.
\eeq
The absolute square of this equation contains three interference terms.
However, adding to that expression a suitable linear combination of
the absolute square of the previous two equations (the coefficients each
turn out to be 1/32), one finds an expression without interference terms:
\beq
|\bra{1',0} H_W \ket{D^0}|^2 = \frac{1}{2}|A(D^0 \to \eta \eta)|^2
+ |A(D^0 \to \eta \pi^0)|^2 + \frac{1}{2}|A(D^0 \to \pi^0 \pi^0)|^2~.
\eeq
When calculating decay rates involving identical particles, one must multiply
the first and last terms by 2, leading to the result
\beq
y(\Delta S = 0) = {\cal B}(D^0 \to \eta \eta) + {\cal B}(D^0 \to \eta \pi^0)
+ {\cal B}(D^0 \to \pi^0 \pi^0)~.
\eeq
The final result for $y(\pi^0, \eta, K^0, \bar K^0)$ is obtained by adding 
this contribution to that from the $\Delta S = \pm 1$ transitions to pairs of 
neutral mesons:
\bea
& & y(\pi^0, \eta, K^0, \bar K^0) = {\cal B}(D^0 \to \eta \eta) + {\cal B}(D^0 \to \eta \pi^0)
+ {\cal B}(D^0 \to \pi^0 \pi^0) \nonumber \\
& &~~~~ -2 \sqrt{[{\cal B}(D^0 \to K^0 \eta) + {\cal B}(D^0 \to K^0
\pi^0)] [{\cal B}(D^0 \to \bar K^0 \eta) + {\cal B}(D^0 \to \bar K^0 \pi^0)]}~.
\label{eqn:neuts}
\eea
The relation $y(\pi^0, \eta, K^0, \bar K^0) = 0$ which holds in USFB
is, of course, satisfied by less precise SU(3) rate relations summarized in 
Table \ref{tab:neuts}. Early examples of  SU(3) analyses may be found 
in Refs.\ \cite{Wang:1979} and \cite{Quigg:1980}. 
The results in Table \ref{tab:neuts} follow from  the 
behavior of the $\Delta S=\pm1, 0$ pieces in $H_W$ as three 
components $U_3=\pm 1,0$ of a U-spin triplet operator.
  
\begin{table}
\caption{Absolute squares of amplitudes $A(D^0 \to f)$ for final states
consisting of two neutral pseudoscalar mesons.  A factor of two has been
included for final states with two identical particles.  An overall common
factor has been omitted.  The $\eta$ is taken as a pure octet member.
\label{tab:neuts}}
\begin{center}
\begin{tabular}{c c} \hline \hline
Final state $f$  &   $|A|^2$   \\ \hline
$\bar K^0 \pi^0$ &  $(1/2)C^4$ \\
$\bar K^0 \eta$  &  $(1/6)C^4$ \\
$\pi^0 \pi^0$ & $(1/2)C^2 S^2$ \\
$\pi^0 \eta$  & $(1/3)C^2 S^2$ \\
$\eta \eta$   & $(1/2)C^2 S^2$ \\
$K^0 \pi^0$      &  $(1/2)S^4$ \\
$K^0 \eta$       &  $(1/6)S^4$ \\ \hline \hline
\end{tabular}
\end{center}
\end{table}

As in the example of charged pions and kaons, first order SU(3)-breaking
contributions from phase space cancel in $\Delta y(\pi^0, \eta, K^0,
\bar K^0)$.  Here we use the rate relations of Table \ref{tab:neuts}. An $\eta$
in the final state counts for 4/3 of a strange quark, as $\eta_8$ is an
$s \bar s$ pair 2/3 of the time.  [This is equivalent to using a Gell-Mann
-- Okubo mass formula (either linear or quadratic) for $M_\eta$ in terms of
$M_K$ and $M_\pi$.]  The contributions to the sum rule (\ref{eqn:neuts}) are
then (neglecting common factors)
\beq
\frac{1}{2} + \frac{1}{3} + \frac{1}{2} - \frac{4}{3} = 0~,
\eeq
while the coefficients of $\Delta$ from these corresponding terms are
\beq
\frac{1}{2} \cdot \frac{8}{3} + \frac{1}{3} \cdot \frac{4}{3} + \frac{1}{2}
\cdot 0 - 2 \cdot \left\{ \frac{1}{2} + \frac{1}{6} \cdot \left[ 1 + \frac{4}{3}
\right] \right\} = \frac{16}{9} - \frac{16}{9} = 0~.
\eeq

No information is available for the decays $D^0 \to K^0 \eta^0$ and $D^0 \to
K^0 \pi^0$, so we can't tell how well Eq.\ (\ref{eqn:neuts}) cancels.  
Since we expect it vanishes to first order in U-spin breaking, we have a sum rule
that may be used to predict the sum of these two DCS branching fractions:
\beq \label{eqn:DCSpred}
{\cal B}(D^0 \to K^0 \eta) + {\cal B}(D^0 \to K^0 \pi^0) = (7.4 \pm 1.2) \times 10^{-5}~.
\eeq
This is for a pure octet $\eta$, but we have argued that $\eta$--$\eta'$ mixing
is second order in SU(3) breaking.  When data become available it will be
interesting to compare this prediction
with  the data and with the central value obtained 
from an SU(3) fit with an $11.7^\circ$ $\eta$--$\eta'$ mixing angle 
\cite{Bhattacharya:2010},
\beq
{\cal B}(D^0 \to K^0 \eta) + {\cal B}(D^0 \to K^0 \pi^0) = (2.8 + 6.9) \times 10^{-5}
= 9.7 \times 10^{-5}~.
\eeq
This value involves however an uncertainty from first order SU(3) breaking
corrections which do not affect the prediction  (\ref{eqn:DCSpred}).

\subsection{Decays to charged PV states}

When one of the final state mesons is a pseudoscalar meson P and the 
other a vector meson V there are more 
U-spin [or SU(3)] amplitudes as the final-state
particles do not belong to the same multiplet.  The U-spin doublets are:
\beq
{\rm Pseudoscalar~mesons:} \left( \begin{array}{c} K^+ \cr -\pi^+\end{array}
\right)~;~~\left( \begin{array}{c} \pi^- \cr K^- \end{array} \right)~;
\eeq
\beq
{\rm Vector~mesons:} \left( \begin{array}{c} K^{*+} \cr -\rho^+ \end{array}
\right)~;~~\left( \begin{array}{c} \rho^- \cr K^{*-} \end{array} \right)~.
\eeq
One can then form U-spin triplet PV states of charge zero out of the above 
doublets in two different ways, using the two pairs $(\pi^-,K^-)$ and
$(K^{*+},\rho^+)$ on the one hand and their charge-conjugates on the other.
A test of the very generic sum rule (\ref{USR}) is quite challenging, as it 
requires measuring relative phases between $D^0$ decay amplitudes for a 
PV state and its charge-conjugate. In principle, this may be achieved by a Dalitz plot
analysis for decays to a common three-body final state. 

Facing this experimental difficulty, we will now study testable U-spin sum
rules similar to (\ref{eqn:chpk}), in which first order U-spin breaking
corrections cancel in phase space factors but may occur in hadronic amplitudes.  
In the U-spin symmetry limit there are two classes of
amplitude relations, depending on which pair of U-spin doublets we consider:
\beq\label{Uhierarchy}
A(D^0 \to \pi^- K^{*+}) = - \lambda A(D^0 \to K^- K^{*+}) = \lambda
A(D^0 \to \pi^- \rho^+) = - \lambda^2 A(D^0 \to K^- \rho^+)~;
\eeq
\beq
A(D^0 \to \rho^- K^+) = - \lambda A(D^0 \to K^{*-} K^+) = \lambda
A(D^0 \to \rho^- \pi^+) = - \lambda^2 A(D^0 \to K^{*-} \pi^+)~,
\eeq
where $\lambda \equiv \tan \theta_c$.  One can form sets of contributions to
$y$ out of either set, but in neither case do we have assurance that
first-order U-spin-breaking effects in hadronic amplitudes cancel one another.
 
\subsubsection{$(K^{*+},\rho^+)(\pi^-,K^-)$ final states}

Using one pair of U-spin doublets, a set of contributions to $y$ for which
branching fractions are known for all four processes is
\beq \label{eqn:TpEv}
y_1 \equiv {\cal B}(D^0 \to \pi^-\rho^+) + {\cal B}(D^0 \to K^-K^{*+}) -
2\sqrt{{\cal B}(D^0 \to K^-\rho^+) {\cal B}(D^0 \to \pi^-K^{*+})} = 0~.
\eeq
Substituting the known branching fractions \cite{PDG}
\bea
{\cal B}(D^0 \to \pi^-\rho^+) & = & (9.8 \pm 0.4) \times 10^{-3}~,\\
{\cal B}(D^0 \to K^-K^{*+}) & = & (4.38 \pm 0.21) \times 10^{-3}~,\\
{\cal B}(D^0 \to K^-\rho^+) & = & (10.8 \pm 0.7) \% ~,\\
{\cal B}(D^0 \to \pi^-K^{*+}) & = & (3.39^{+1.80}_{-1.02}) \times 10^{-4}~,
\eea
one finds $y_1 = (2.1^{+1.9}_{-3.3})\times 10^{-3}$, with the error dominated
by the uncertainty in the last branching fraction.  Some cancellation occurs,
but it is not as well-determined as for charged pions and kaons (Sec.\ VI A).

The effects of U-spin breaking in phase space factors lead to first-order
corrections proportional to $M_K^2 - M_\pi^2$ or $M_{K^*}^2 - M_\rho^2$,
both of which can be seen to cancel one another in Eq.\ (\ref{eqn:TpEv}).

\subsubsection{$(K^+,\pi^+)(\rho^-,K^{*-})$ final states}

Using the other combination of P and V U-spin doublets, one can write their
contribution to $y$ as
\beq \label{eqn:TvEp}
y_2 \equiv {\cal B}(D^0 \to \rho^-\pi^+) + {\cal B}(D^0 \to K^{*-}K^+) -
2\sqrt{{\cal B}(D^0 \to K^{*-}\pi^+) {\cal B}(D^0 \to \rho^-K^+)} = 0~.
\eeq
We have almost enough information to check this sum rule:
\bea
{\cal B}(D^0 \to \rho^-\pi^+) & = & (4.96 \pm 0.24) \times 10^{-3}~,\\
{\cal B}(D^0 \to K^{*-}K^+) & = & (1.56 \pm 0.12) \times 10^{-3}~,\\
{\cal B}(D^0 \to K^{*-}\pi^+) & = & (5.63 \pm 0.35)\%~,\\
{\cal B}(D^0 \to \pi^- \pi^0 K^+) & = & (3.04 \pm 0.17) \times 10^{-4}~.
\eea
The value ${\cal B}(D^0 \to K^{*-} \pi^+) = (5.65 \pm 0.35)\%$ quoted above is three
times the average of the values \cite{PDG} ${\cal B}(D^0 \to K^{*-} \pi^+ \to K^-
\pi^0 \pi^+) = (2.22^{+0.40}_{-0.19})\%$ and ${\cal B}(D^0 \to K^{*-} \pi^+ \to K_S
\pi^- \pi^+) = (1.66^{+0.15}_{-0.17})\%$, using the lower error bar for the
first and the upper error bar for the second (because the average lies
between them).

The most recent data contributing to this last branching fraction are
from Belle \cite{Tian05} and BaBar \cite{Aubert06N}.  The former makes no
statement about how much of the $\pi^- \pi^0$ state corresponds to a
$\rho^-$, but a $\rho^-$ is clearly visible in the Dalitz plot of the
latter.  Assuming that all the $\pi^- \pi^0$ is in a $\rho^-$, one obtains
a value of $y_2 = (-1.75 \pm 0.44) \times 10^{-3}$, but one may be
oversubtracting.  It would be very useful if an analysis of the decay
$D^0 \to \pi^- \pi^0 K^+$ could extract ${\cal B}(D^0 \to \rho^- K^+)$.

As in the case of $y_1$, the contributions of SU(3) breaking in the phase space
factors of $y_2$ cancel one another to first order.

\subsubsection{Using both pairs of U-spin multiplets}

One can write a sum rule involving all eight PV modes which involves only
the products of decay amplitudes and their charge conjugates.  In the absence
of strong phase differences one then finds a contribution to $y$ of the form
\bea \nonumber
   y_3 &=& \sqrt{{\cal B}(D^0 \to \rho^- \pi^+){\cal B}(D^0 \to \rho^+\pi^-)}
        +  \sqrt{{\cal B}(D^0 \to K^{*-} K^+) {\cal B}(D^0 \to K^{*+}K^-)} \\
       &-& \sqrt{{\cal B}(D^0 \to K^{*-}\pi^+){\cal B}(D^0 \to K^{*+}\pi^-)}
        -  \sqrt{{\cal B}(D^0 \to \rho^+ K^-) {\cal B}(D^0 \to \rho^-K^+)}
\eea

Evaluation yields $y_3 = (-0.5^{+0.8}_{-1.2}) \times 10^{-3}$.
This is reassuringly small, but we have not justified the neglect of the
strong phase differences between the amplitudes 
for charge-conjugate final states, contributing to $y_1$
(proportional to $T_P + E_V$ in the notation of Ref.\ \cite{Bhattacharya:PV})
and to $y_2$ (proportional to $T_V + E_P$ in that notation).  An analysis
of related charm decays to $PV$ final states finds a small but non-negligible
strong phase difference between the two \cite{Bhattacharya:PV,Bh}.

\section{Four-body states of charged pions and kaons}

The states of four pions and/or kaons were identified in Ref.\
\cite{Falk:2001hx} as likely candidates for substantial SU(3) breaking in
$D^0$--$\od^0$ mixing.  The four-kaon channel is closed to $D^0$ decays
so arguments based on the cancellation of first-order SU(3) breaking or 
U-spin-breaking effects will fail. 

A full analysis of cancellations in four-body final states would require
comparison of similar kinematic regions for individual U-spin multiplets.
This is beyond the scope of the present article, but we can identify some
useful groupings of charge pions and kaons.  These belong to U-spin doublets,
as mentioned earlier, so the U-spin multiplets containing them are those
in the product
\beq
\left[ U=\frac{1}{2} \right] ^4 = 1(U=2) + 3(U=1) + 2(U=0)~.
\eeq
It is the $U=1$ multiplets which interest us as they are 
the only ones reached from
$D^0$ via $H_W^{\Delta C = -1}$.  Three mutually
orthogonal $U=1$ multiplets are summarized in Table \ref{tab:usp}.

\begin{table}
\caption{U-spin triplets of four charged pions and kaons.
\label{tab:usp}}
\begin{center}
\begin{tabular}{c c c} \hline \hline
Multiplet & Norm & Meson states \\ \hline
$\ket{1,1}_1$ & 1/2    & $-\ket{\pi^+ K^+\pi^-\pi^-}+\ket{K^+\pi^+\pi^-\pi^-}
  +\ket{K^+K^+K^-\pi^-} - \ket{K^+K^+\pi^-K^-}$ \\
$\ket{1,0}_1$ & $1/\s$ & $-\ket{\pi^+K^+K^-\pi^-} + \ket{K^+\pi^+\pi^-K^-}$ \\
$\ket{1,-\!1}_1$ & 1/2 & $\ket{\pi^+\pi^+K^-\pi^-}-\ket{\pi^+K^+K^-K^-}
  -\ket{\pi^+\pi^+\pi^-K^-}+\ket{K^+\pi^+K^-K^-}$ \\ \hline
$\ket{1,1}_2$ & 1/2    & $-\ket{\pi^+ K^+\pi^-\pi^-}-\ket{K^+\pi^+\pi^-\pi^-}
  -\ket{K^+K^+K^-\pi^-} - \ket{K^+K^+\pi^-K^-}$ \\
$\ket{1,0}_2$ & $1/\s$ & $\ket{\pi^+\pi^+\pi^-\pi^-}-\ket{K^+K^+K^-K^-}$ \\
$\ket{1,-\!1}_2$ & 1/2 & $\ket{\pi^+\pi^+K^-\pi^-}+\ket{\pi^+\pi^+\pi^-K^-}
  +\ket{\pi^+K^+K^-K^-} + \ket{K^+\pi^+K^-K^-}$ \\ \hline
$\ket{1,1}_3$ & 1/2    & $-\ket{\pi^+ K^+\pi^-\pi^-}+\ket{K^+\pi^+\pi^-\pi^-}
  -\ket{K^+K^+K^-\pi^-} + \ket{K^+K^+\pi^-K^-}$\\
$\ket{1,0}_3$ & $1/\s$ & $-\ket{\pi^+K^+\pi^-K^-}+\ket{K^+\pi^+K^-\pi^-}$ \\
$\ket{1,-\!1}_3$ & 1/2 & $\ket{\pi^+\pi^+\pi^-K^-}-\ket{\pi^+K^+K^-K^-}
  -\ket{\pi^+\pi^+K^-\pi^-}+\ket{K^+\pi^+K^-K^-}$ \\ \hline \hline
\end{tabular}
\end{center}
\end{table}

If they obey the pattern of previous examples, the sum rules will involve
cancellations of $U_3 = 0$ contributions against ones of $U_3 = \pm1$.
By counting kaons one can see that, at  least formally, first order 
U-spin-breaking corrections in phase space seem to cancel one another.
However, the $\ket{1,0}_2$ state
is particularly susceptible to U-spin-breaking because it is the only one which
contains the state of four charged kaons.  Hence if a source of a significant
contribution to $y$ is to be sought in the states of four kaons, the
sum rule associated with the triplet $\ket{1,U_3}_2$ would be a good place
to look. As we will show now this contribution is expected to be negative, while
the measured value of $y$ is positive~\cite{HFAG}.
 
One may assume that nonresonant four-body decays are dominated by states in
which relative angular momenta for all particle pairs are zero, so that the
state with four charged pions is CP-even. The branching fraction for a nonresonant
state involving three charged kaons and a charged pion~\cite{PDG}, 
${\cal B}(K^+K^-K^-\pi^+)_{\rm nonres}= (3.3\pm 1.5)\times 10^{-5}$,
is three orders of magnitude smaller than the branching fraction for a single kaon and 
three pions~\cite{PDG}, ${\cal B}(K^-\pi^-\pi^+\pi^+)_{\rm nonres}$ 
$= (1.88 \pm 0.26)\%$, and may be neglected. Thus the contribution to $y$ from 
the triplet $\ket{1,U_3}_2$ is given by an expression similar to (\ref{eqn:chpk}),
but a term ${\cal B}(K^+K^+K^-K^-)$ is missing on the right-hand-side,

\beq
y(\ket{1,U_3}_2) = {\cal B}(\pi^+\pi^+\pi^-\pi^-) - 
2\sqrt{{\cal B}(K^-\pi^-\pi^+\pi^+){\cal B}(K^+\pi^+\pi^-\pi^-)}~.
\eeq
All three branching ratios correspond to nonresonant four-body final states. Assuming 
the usual U-spin hierarchy 
similar to (\ref{Uhierarchy}),
\beq
{\cal B}(K^+\pi^+\pi^-\pi^-):{\cal B}(\pi^+\pi^+\pi^-\pi^-):{\cal B}(K^-\pi^-\pi^+\pi^+)
\simeq \lambda^4:\lambda^2:1~,
\eeq
one finds $y(\ket{1,U_3}_2)$ to be negative. 
Using the measured value of
${\cal B}(K^-\pi^-\pi^+\pi^+)_{\rm nonres}$ to normalize the other two branching 
fractions one obtains $y(\ket{1,U_3}_2) \simeq - 1.0\times 10^{-3}$.

One may wonder whether a U-spin sum rule exists also for four-body $D^0$ decays 
involving $K^0 \bar K^0$ and a pair of charged pions or kaons.
A U-spin relation following from a symmetry under a $d \leftrightarrow s$ reflection,
\beq
\langle K^0\bar K^0\pi^+\pi^-|H_W|D^0\rangle = - \langle \bar K^0 K^0 K^+ K^-|H_W|D^0
\rangle~,
\eeq
is strongly broken by phase space which forbids decays into four kaons~\cite{PDG},
\beq
{\cal B}(D^0\to K^0\bar K^0 \pi^+\pi^-) = (4.92 \pm 0.92)\times 10^{-3}~,~~~~~~~
{\cal B}(D^0 \to \bar K^0 K^0 K^+ K^-) = 0~.
\eeq
The first branching ratio would explain the measured 
value of $y$, if a sum rule including the difference
\beq\label{diff}
{\cal B}(K^0 \bar K^0\pi^+\pi^-) - 
2\sqrt{{\cal B}(K^0 \bar K^= \pi^+ K^-){\cal B}(K^0 \bar K^0 K^+\pi^-)}
\eeq
could be obtained for a U-spin triplet state.
$K^0$ and $\bar K^0$ are two members of a U-spin triplet to which also  
$(\sqrt3\eta-\pi^0)/\s$ belongs. [See Eq.\,(\ref{singleU=1})]. One may 
show that in fact there exists a U-spin triplet sum rule including the difference (\ref{diff})
which, however,  involves also unmeasured branching fractions and interference terms 
with amplitudes involving $\pi^0$ and $\eta$ in addition to a pair of charged pions or kaons.

\section{Conclusions}

In the limit of U-spin symmetry, contributions of on-shell intermediate
states to the parameter $y = \Delta \Gamma/(2 \Gamma)$ describing 
$D^0$--$\od^0$ mixing cancel one another.  This has been shown to be a
consequence of the fact that the mixing amplitude transforms as a U-spin
operator with $U=2$, $U_3 = 0$, while the states $\ket{D^0}$ and $\ket{\od^0}$
have $U=0$ because they contain no $s$ or $d$ quarks or antiquarks.

The cancellation of first-order U-spin-breaking effects then follows from
the fact that first-order U-spin-breaking (equivalent to insertion of a
term $m_s-m_d$) transforms as $U=1,U_3=0$ and therefore cannot contribute
to the mixing.

This result implies that sum rules may be written for contributions to $y$
each involving a distinct U-spin triplet, explaining, for example, why
the cancellation of contributions from the intermediate states $K^- \pi^+,
K^-K^+,\pi^-\pi^+$, and $K^+ \pi^-$ occurs.  These states belong
to a $U=1$ multiplet all of whose members are sufficiently far below
$M_D$ that U-spin-breaking effects in phase space factors may be treated to
first order in perturbation theory, and indeed -- as expected from the general
theorem -- they cancel one another to first order in U-spin-breaking.

Examples of multiplets for which cancellation of contributions to $y$ cancel
one another have been given.  In addition to the above case of pairs of
charged kaons or pions, sum rules are seen to hold for pairs of neutral
members of the pseudoscalar octet, pairs of charged pseudoscalar and vector
mesons, and specific groupings of four charged pions and kaons.

When looking for standard model culprits which could induce large values of
$y$, Ref.\ \cite{Falk:2001hx} identified final states consisting of four
particles, noting that four-kaon final states lie above $M_D$ and hence are
inaccessible.  Perturbative U-spin-breaking is thus a very poor approximation
for sum rules involving such states.  We have identified a grouping of
amplitudes which includes the state $\ket{K^+K^+K^-K^-}$ and thus is a
good candidate to participate in strong U-spin-breaking contributions to $y$.
We have shown that this contribution is most likely negative of order $-10^{-3}$, 
while the value measured for $y$ is positive at a level of an appreciable fraction
of a percent. 

\section*{Acknowledgments}

We thank the CERN Theory Group for hospitality during part of this work.
J.L.R. was supported in part by the United States Department of Energy
under Grant No.\ DE-FG02-90ER40560.

\end{document}